\documentclass[twocolumn]{aastex63} 

\usepackage{amsmath}
\usepackage{amssymb}
\usepackage{empheq}
\usepackage{mathrsfs}
				
\usepackage{amssymb}

\received{\today}
\revised{\today}
\accepted{\today}
\submitjournal{ApJL}

\shorttitle{Timing of the Giant Planet Instability}
\shortauthors{Morgan, Seligman, Batygin}
\graphicspath{{./}{figures/}}
\begin{document}

\title{Collisional Growth Within the Solar System's Primordial Planetesimal Disk and the Timing of the Giant Planet Instability}

\correspondingauthor{Marvin Morgan}
\email{marv08@sas.upenn.edu }

\author[0000-0003-4022-6234]{Marvin Morgan}
\affiliation{University of Pennsylvania Department of Physics and Astronomy, Philadelphia, PA 19104, USA}

\author[0000-0002-0726-6480]{Darryl Seligman}
\affiliation{Dept. of the Geophysical Sciences, University of Chicago, Chicago, IL 60637}

\author[0000-0002-7094-7908]{Konstantin Batygin}
\affil{Division of Geological and Planetary Sciences, Caltech, Pasadena, CA 91125}




\begin{abstract}
The large scale structure of the Solar System has been shaped by a transient dynamical instability that may have been triggered by the interaction of the giants planets with a massive primordial disk of icy debris. In this work, we investigate the conditions under which this primordial disk could have coalesced into planets using analytic and numerical calculations. In particular, we perform numerical simulations  of the Solar System's early dynamical evolution that account for the  viscous stirring and collisional damping within the disk. We demonstrate that if collisional damping would have been sufficient to maintain a temperate velocity dispersion, Earth mass trans-Neptunian planets could have emerged within a timescale of 10 Myr. Therefore, our results favor a scenario wherein the   dynamical instability of the outer Solar System began immediately upon the dissipation of the gaseous nebula to avoid the overproduction of Earth mass planets in the outer Solar System. 

\end{abstract}


\section{Introduction} \label{sec:intro}

The emergent picture of the evolution of the early Solar System is an instability-driven scenario.  \citet{Tsiganis2005} proposed what is commonly refered to as the ``Nice-model,'' in which the giant planets formed on circular, coplanar orbits with an accompanying planetesimal disk  located between $\sim 15 - 35$au.  Subsequent interactions with this primordial disk triggered a dynamical instability. 

The Nice model  reproduces a variety of characteristics of the present day Solar System, including the current orbits of the giant planets, the  inclination distribution of the co-orbital Jupiter  Trojans \citep{Morbidelli2005,Nesvorny2013} and the existence and structure of the Kuiper belt \citep{Levison2008,Nesvorny2012,Nesvorny2015,gomes2018}. For a recent review of the early evolution of the Solar System, we refer the reader to \citet{Nesvorny2018}.

Although the Nice model  successfully  accounts for the large-scale structure of the Solar System, numerous details remain elusive. Particularly, the precise timing of the instability is somewhat unconstrained. A promising avenue  to constrain the timing of the instability was to relate it to the Late Heavy Bombardment  (LHB) of the moon. The petrological record  of Lunar craters implies that the Moon experienced a bombardment flux of planetesimals, roughly $\sim 700$ Myrs after  the planets formed \citep{Hartmann2000}. It is uncertain whether this bombardment  occurred during a cataclysmic spike or one that slowly decayed over time. This increase in cratering events could be caused by an increase of the bombardment rate \citep{Tera1974,Ryder1990,Ryder2002}. \citet{Gomes2005} demonstrated that the migration of the giant planets would naturally produce a sudden flux of planetsimals into the inner Solar System, which would explain the LHB. \citet{Levison2011}, however, demonstrated that the exchange of energy between the giant planets and the planetesimal disk
would only explain the timing of the LHB if the disk's inner edge was sufficiently removed from the orbit of Neptune.   Alternatively, the LHB could also be explained by impacts from left over planetesimals at the  tail-end of terrestrial planet accretion \citep{Hartmann1975,Neukum2001,Hartmann2003}.


In a set of recent studies, numerous  authors have proposed that the Nice model instability and the LHB are unrelated and have argued for  an early instability that started $<100$ Myrs after the formation of the Sun \citep{deSousa2019,Nesvorny2021}. \citet{Nesvorny2018b}  argued that the existence of the  binary Jupiter Trojan (617) Patroclus–Menoetius \citep{Grav2011,Buie2015} implied that the primordial disk dissipated via migrating planets within $\sim100$ Myr of the formation of the Sun. Beyond these concerns, aspects of the terrestrial planets are more readily reproduced with the early instability scenario. In particular, the survival of the terrestrial planets is  more likely in the early instability model \citep{Clement2018,Clement2019}. 


In this work, we study a related, but  distinct aspect of the instability driven scenario: the evolution of the planetesimal disk. For computational purposes, most studies ignore self-gravity — and thus the possibility of growth generally — in the planetesimal swarm. Accordingly, the possibility that the planetesimal disk could coalesce into planets has not been explored in detail. In this work, we ask a simple question: if the Sun was encircled by $\sim20 {\rm M}_\oplus$ of debris for millions of years, could these debris have formed planets? We answer this question through analytic estimates and direct N-body simulations.  The remainder of this paper is organized as follows. In \S \ref{sec:analytic}, we consider estimates of planetesimal growth rates and the excitation of their velocity dispersion from analytic grounds. In \S \ref{sec:numerical}, we report the results of our numerical experiments and discuss the implications of our results in \S \ref{sec:discussion}.

\begin{figure*}
\begin{center}
\includegraphics[scale=0.85,angle=0]{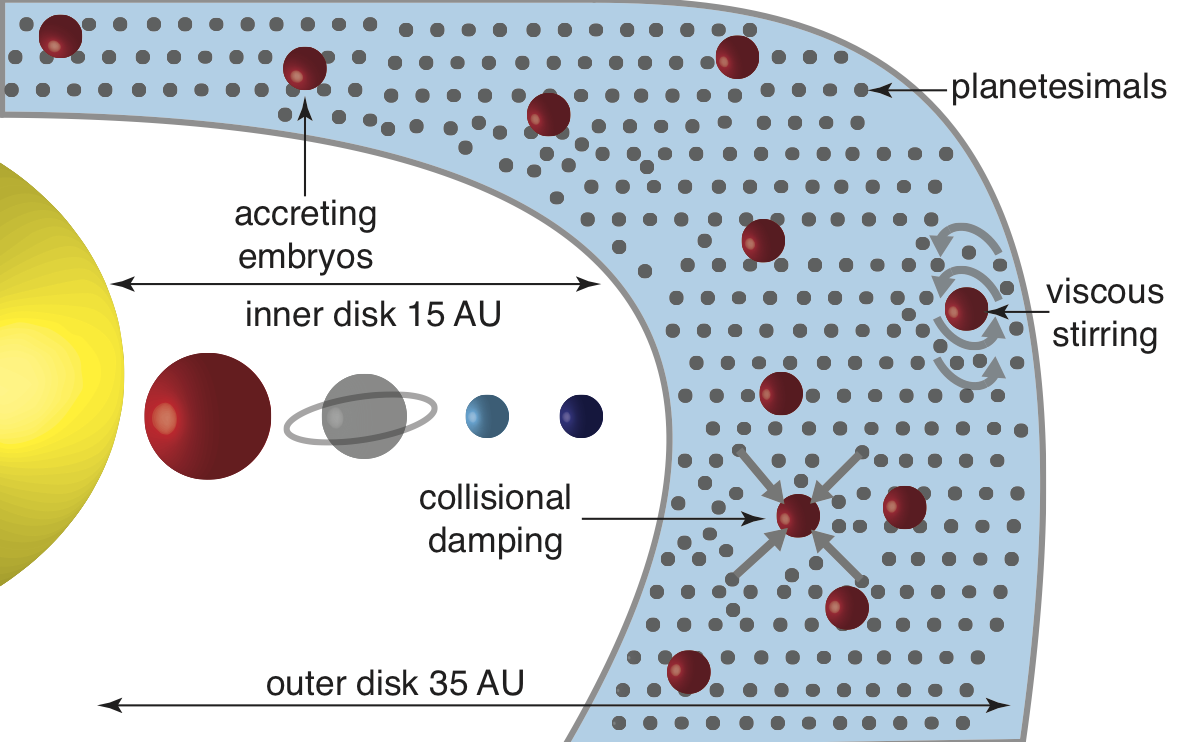}
\caption{A schematic diagram depicting the initial conditions and components of the numerical simulations in \S \ref{sec:numerical}. The integrations have the four giant planets at their primordial locations. The primordial disk is composed of passive planetesimals and accreting embryos, and is located between 15-35 au. The effects of viscous stirring and collisional damping, discussed in \S \ref{sec:analytic}, are implemented throughout the simulation. }\label{Fig:schematic}
\end{center}
\end{figure*}

\section{Analytic Estimates}\label{sec:analytic}
In this section, we consider accretion within the Solar System’s planetesimal disk from analytic grounds. In particular, the goal of the following analysis is to obtain an  estimate for the timescale of mass-growth that is expected to unfold in the primordial trans-Neptunian region of the Solar System, and highlight its connection to the velocity dispersion.

The characteristics of the debris belt that drives planetary migration in the Nice model are readily summarized. The disk is typically assumed to be comprised of solids with a  mass, $M_{\rm{disk}}\sim20\,M_{\oplus}$, and to extend from $r_{0}\sim15\,$au to $r_{\rm{out}}\sim35\,$au. Moreover, the disk is envisioned to emerge from the Solar nebula in a dynamically cold state, meaning that the planetesimals that comprise this belt originate on roughly circular and co-planar orbits. For simplicity, here we adopt the  surface density profile proposed in \citet{Mestel1963}, with $\Sigma=\Sigma_0(r_0/r)$, where $\Sigma_0$ is the planetesimal surface density at the disk's inner edge and $r$ is the radial distance. For simplicity, we assume that the velocity dispersion of the planetesimal belt is initially considerably smaller than the planetesimals' escape velocity, such that the Safronov number, $\Theta=(v_{\rm{esc}}^2/\langle v \rangle^2)$, satisfies $\Theta \gg 1$. Here, $v_{\rm{esc}}$ and $\langle v \rangle$ are the escape velocity and velocity dispersion of the planetsimals. The latter assumption is optimistic, and implies that the initial growth is predominantly facilitated by gravitational focusing \citep{Safronov1972}.

Generically speaking, planetesimal growth can proceed via pebble accretion \citep{Ormel2010,Lambrechts2012}  or through pairwise collisions \citep{Lissauer1993}. Although the former process is nominally faster, it is facilitated by the existence of the gaseous disk and does not operate in the absence of hydrodynamical drag forces. In this work, we are specifically concerned with post-nebular growth of planetesimals. Therefore, we   ignore the effects of pebble accretion altogether and focus entirely on   mutual impact driven assembly. The collisional mass-accretion rate of planetesimals, $\dot{M}$,  can be  estimated via an $n-\sigma-\langle v \rangle$  relation. 

The product of the typical planetesimal mass, $m$, and number density, $n$ is given by $m\,n\sim\Sigma/h$, where $h$ is the characteristic scale-height of the planetesimal disk. The velocity dispersion can be written as $\langle v \rangle\sim v_{\rm{kep}}(r)\,(h/r)$, where $v_{\rm{kep}}(r)$ is the Keplerian velocity at a given radial distance, $r$. The characteristic accretion rate, $\dot{M}$, of a planetesimal is given by, 
\begin{equation}
\dot{M}= 4\,\pi\,\rho\,R^2\,\frac{dR}{dt}\sim\Sigma\,\pi\,R^2\,(1+\Theta)\,\bigg(\frac{v_{\rm{kep}}(r)}{r}\bigg)\, ,
\label{R}
\end{equation}
where $\rho$ is the density of the planetesimal. The gravitational focusing is accounted for by the $(1+\Theta)$ enhancement of the collisional cross-section. The derivation of Equation \ref{R}  is analogous to the formalism developed in \citet{Armitage2010}. This expression yields a constant $dR/dt$, because  $M\propto t^3$. This  can be recast as the characteristic timescale, $\tau_{\rm{accr}}$, that is necessary for the radius to increase by some value $\Delta R$,
\begin{equation}
\tau_{\rm{accr}}\sim\frac{4\,\rho\,\Delta R}{\Sigma\,\Omega\,(1+\Theta)}.
\label{Taccr}
\end{equation}
We can estimate the value of the surface density using $M_{\rm{disk}}\sim2\,\pi\,r_0\,\Sigma_0\,(r_{\rm{out}}-r_0)$, where $r_0$ and $r_{\rm{out}}$ are the inner and outer disk radii respectively. We obtain $\tau_{\rm{accr}}\sim10^{10}/\Theta$ years for $\Delta R\sim10^3\,$km, $\rho\sim2\,$g/cc and $\Theta\gg1$. 

Equation (\ref{Taccr}) highlights the fact that significant growth within the planetesimal disk can take place on a $\sim100\,$Myr timescale \textit{only} if the disk remains dynamically cold throughout this time. The ratio of the escape velocity to orbital velocity is on the order of $v_{\rm{esc}}/v_{\rm{kep}}\sim0.01$ for an $R\sim100\,$km object orbiting at $r_0$. Therefore the  estimate above implies that sustained growth requires eccentricities and inclinations smaller than $\sim1\%$. In reality, the velocity dispersion of the planetesimal disk is controlled by an assortment of factors, including self-gravitational (viscous) stirring, collisional damping, and the system's approach towards equipartition. 

To estimate the degree of orbital crossing that is expected to develop within the disk, we consider the competing effects of collisional damping and viscous stirring. For this order of magnitude calculation, we neglect planet-driven evolution and accretion, such that  the disk is comprised of equal-mass bodies. In this regime, equipartition (which drives dynamical cooling of massive objects at the expense of dynamical heating of low-mass planetesimals) is inconsequential and all objects are characterized by a common velocity dispersion.

The evolution of the velocity dispersion is given by \citet{Armitage2010}, and can be estimated as, 
\begin{equation}
\frac{d\langle  v \rangle}{dt} =\frac{2 \, \pi \, G^{2} \, m\, \Sigma \,\Omega \, \ln(\Lambda)}{\langle  v \rangle^{3}} - \frac{\langle  v \rangle}{t_{\rm{col}}},
\label{eqnevol}
\end{equation}
where $m$ is the planetesimal mass, and $\ln(\Lambda)\sim10$ is the Coulomb logarithm. As above, the collisional timescale is given by the $n-\sigma-\langle v \rangle$ relation, and takes the form,
\begin{equation}
t_{\rm{col}}=\frac{1}{n\,\pi\,R^2\,\langle  v \rangle}=\frac{2\,m}{\pi\,R^2\,\Sigma\,\Omega}.
\end{equation}

In the absence of collisional damping, Equation (\ref{eqnevol}) dictates that the velocity dispersion grows as $\langle v \rangle \propto t^{1/4}$. For finite $t_{\rm{col}}$, however, there exists an equilibrium solution. Noting the approximate relationship, $\langle v \rangle \sim e\,v_{\rm{kep}}$, this equilibrium corresponds to a characteristic eccentricity given by, 

\begin{equation}
\langle e\rangle\sim \bigg( \frac{2\,\pi\,\ln(\Lambda)\,G^2\,m\,\Sigma\,\Omega\,t_{\rm{col}}}{v_{\rm{kep}}^4}  \bigg)^{1/4}\lesssim0.02\,,
\end{equation}
for an  $R\sim100$ km body, which is the typical planetesimal size produced by the streaming instability \citep{Youdin2005,Johansen2007}. This  expression provides a useful gauge for the effective dynamical temperature of the disk.  It is important to note that this estimate does not account for planetary perturbations (which will stir the planetesimal swarm further), or dynamical friction (which will facilitate accretion of larger embryos, as is typical for simulations of oligarchic growth). Generally, these processes cannot be quantified analytically. Since our analytical estimates are conservative  and do not account for gravitational focusing, in the next section, we use numerical simulations to model trans-Neptunian accretion in the primordial Solar System.

\begin{figure*}
\begin{center}
\includegraphics[scale=0.5,angle=0]{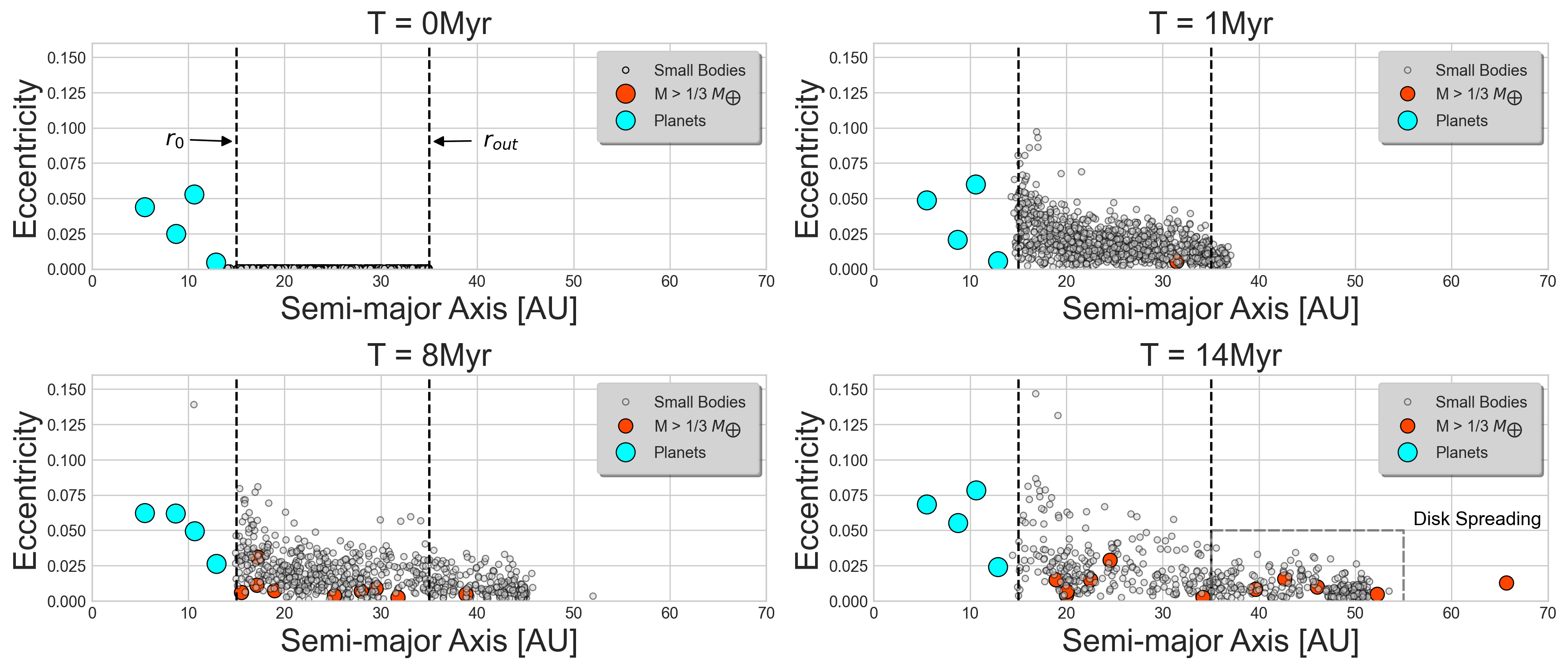}
\caption{The evolution of eccentricities and semi-major axes for all of the objects in the numerical simulation. Small bodies, roughly $\sim1{\rm M}_\oplus$ bodies and giant planets are indicated by grey, red and blue points, respectively. The four panels show the orbital elements at times corresponding to $0,1,8$ and $14$ Myrs in the simulation.  }\label{Fig:eccentricity}
\end{center}
\end{figure*}

\section{Numerical Experiments}\label{sec:numerical}
To simulate accretion within the planetesimal disk, we used the \texttt{mercury6} code  \citep{1999MNRAS.304..793C}. Our numerical experiments follow the standard version of the Nice-model. Planets were initialized in a compact multi-resonant configuration \citep{Batygin2010}, encircled by a $20{\rm M}_\oplus$   disk of planetesimals. The disk's inner and outer edges were set to $15\,$au and $35\,$au, respectively. The calculations were performed using the hybrid Wisdom-Holman/Bulirsch-Stoer algorithm \citep{Wisdom1991}.  The time-step, $\Delta t$ and accuracy parameter, $\epsilon$, were set to $\Delta t=300 $ days and $\epsilon = 10^{-12}$, respectively. 

We modeled the planetesimal disk itself with $N_{\rm{s}}=1000$ super-planetesimals (to represent the larger population of small planetesimals within the disk), and seeded it with $N_{\rm{b}}=100$ protoplanetary embryos (assumed to be fully formed at the start of the simulation and can accrete mass). All planetesimal and embryo masses were set to a common value at the beginning of the simulation, $10^{-8}$ and $10^{-7}$ $M_\odot$ with radii $\sim2000$km and $\sim5200$km for fiducial small and large seeds respectively. All particles  were initialized with negligible eccentricities and inclinations. \citet{Nesvorny2020} presented a methodology  to assign radii to superparticles such  that numerical simulations have the correct collisional timescale. Here, we assume that each simulated protoplanetary embryo  represents a fully formed, spherical particle  with a bulk density of $1{\rm g/cm}^3$. While this is an appropriate approximation for the accreting protoplanetary embryos,  it is not necessarily appropriate for the superparticles, which represent a larger number of planetesimals.   We performed convergence tests   with initial masses of superparticles that were $\sim1/2$ and $\sim2\times$ the inital mass, and verified that the results of the simulations did not sensitively depend on the initial radii of the particles. While the interactions among the embryos and planets were self-consistently modeled in a conventional $N-$body fashion, self-gravitational coupling among the planetesimals was neglected to conserve computational costs. All collisions were treated as perfect mergers.

Despite being a necessary approximation for computational efficiency, the super-particle modeling scheme leads to a strongly over-excited velocity dispersion. This is because a vast population of undamped massive planetesimals generates much stronger gravitational scattering events than those that occured in the real disk. This leads to an unphysical excitation of the distribution of the planetesimals'  eccentricities and inclinations. To counteract this effect in our calculation, we mimicked the effects of collisional damping and dynamical friction by introducing a fictitious acceleration, $\vec{a}_{\rm{damp}}$, into the equations of motion at each timestep \citep{Papaloizou2000} of the form,
\begin{equation}
\vec{a}_{\rm{damp}}= -2 \,\bigg( \frac{\vec{v} \cdot \vec{r}}{r^{2}\,\tau}\bigg) \hat{r} -2\bigg( \frac{\vec{v} \cdot \vec{k}}{\tau}\bigg) \hat{k}     \,,
\end{equation}
where $\vec{k}=(0,0,z)$ is the vector corresponding to the z-component of the planetesimals' position.

\begin{figure*}
\begin{center}
\includegraphics[scale=0.5,angle=0]{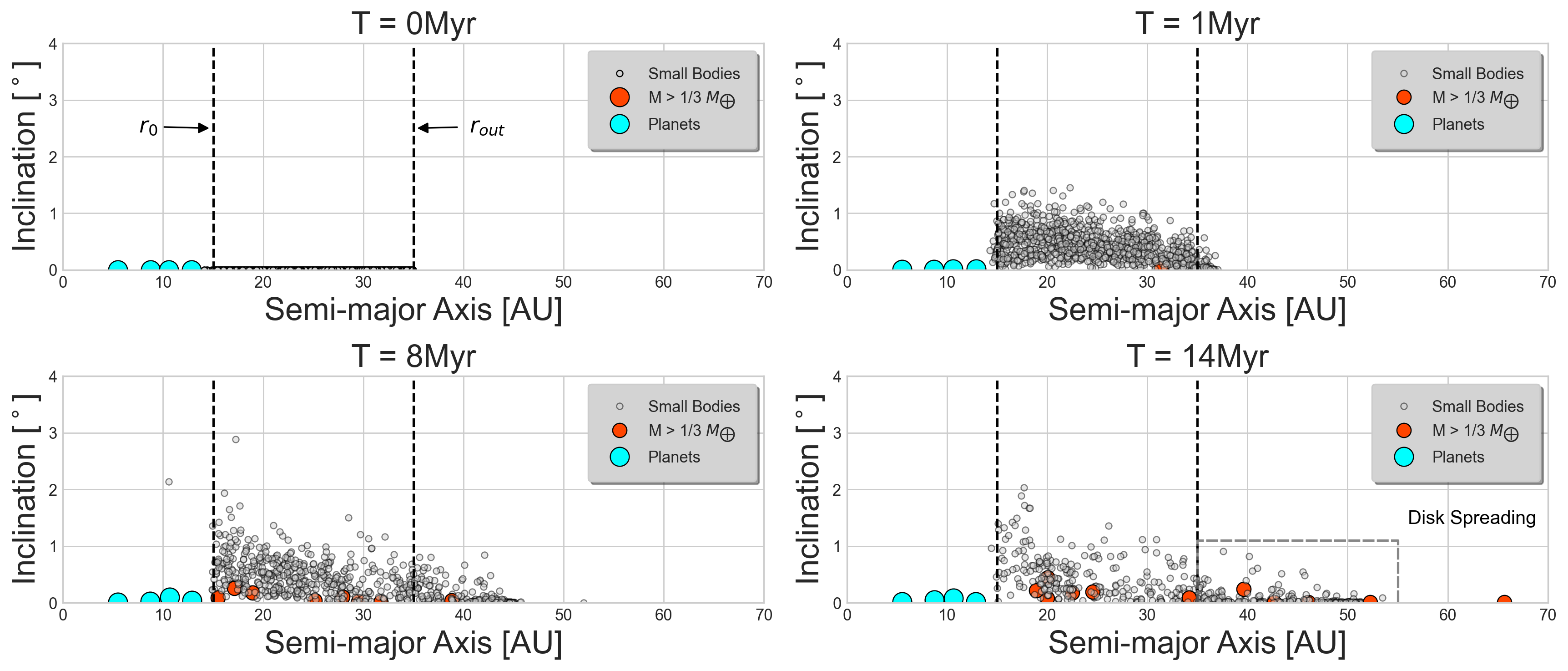}
\caption{Same as Figure \ref{Fig:eccentricity}, for inclinations and semi-major axes. }\label{Fig:inclination}
\end{center}
\end{figure*}

Simply put, this acceleration does not affect objects that occupy circular and planar orbits, but damps the eccentricities and inclinations on a characteristic timescale, $\tau$, if they develop. To calibrate the damping time in our numerical experiments, we tuned the value of $\tau$ until our simulation reproduced the viscous stirring -- collisional damping equilibrium discussed in the previous section. We did this by removing the giant planets from the simulation and suppressing physical collisions within the disk such that the planetesimals and embryos retain equal masses in perpetuity.

Varying $\tau$, we carried out each test integration for $\sim10$ damping times, and recovered the $\langle e \rangle\propto \tau^{1/4}$ relationship predicted by equation (\ref{eqnevol}). Accordingly, we found that a damping time of $\tau=50{,}000\,$years yields $\langle e \rangle = 0.014$ -- a value that is close to our analytic estimate (see Table \ref{tab:tau} and the discussion in the final paragraph of this section). This result is consistent with the numerical results presented by \citet{Levison2011}, who carried out detailed simulations of the early dynamical evolution of the Solar System, and also found that viscous stirring yields root-mean-square disk eccentricity of approximately 2\%. Adjusting the gravitational potential length scale for each particle could provide an alternative solution to damping the over-excitation of particles. 

With our numerical setup calibrated, we restored the giant planets into the simulation and evolved the Solar System, fully accounting for growth of protoplanetary embryos through pairwise impacts. In Figure \ref{Fig:schematic}, we show a schematic of the initial conditions of and relevant processes included in  the simulation.  Intriguingly, our simulations indicate that if the planetary instability or rapid outward migration of Neptune does not ensue shortly after the dissipation of the protosolar nebula, then significant collisional growth of embryos unfolds on a $\sim10\,$Myr timescale. 

\begin{figure}
\begin{center}
\includegraphics[scale=0.2,angle=0]{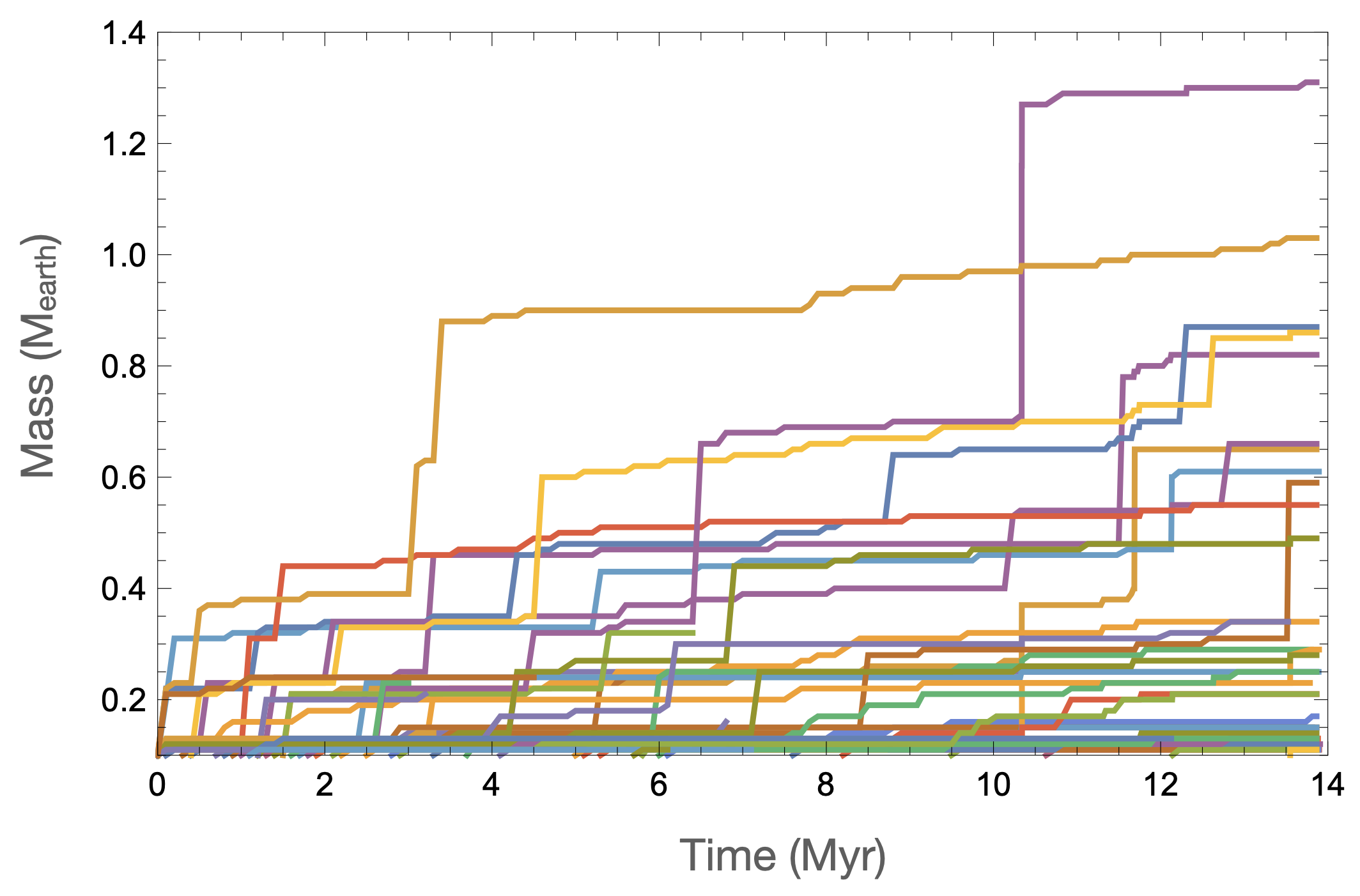}
\caption{The mass evolution of the 100 accreting embryos in the simulation.  The oligarchic growth within the disk occurs on timescales of order $ \tau \sim  10$ Myr. }\label{Fig:mass_evolution}
\end{center}
\end{figure}

A series of snapshots of the dynamical state of the outer Solar System attained within our fiducial simulation are shown in Figure \ref{Fig:eccentricity} and Figure \ref{Fig:inclination}. In some similarity with standard models of terrestrial planet accretion \citep{Chambers2001,Raymond2011}), our numerical experiment shows how the initially circular and planar disk of planetesimals begins to coalesce, and because of dynamical friction, growing embryos settle to the midplane where their continued accretion is further aided by a diminished velocity dispersion. Within $\sim10\,$Myr, numerous $\sim {\rm M}_\oplus$ planets emerge within the disk, and the system eventually enters a slower mode of accretion. The mass evolution of the  accreting embryos  within our model disk is shown in Figure \ref{Fig:mass_evolution}. Of the 100 initial  embryos in the simulation, 2 grow to larger than $\sim1{\rm M}_\oplus$ and $\sim 10\%$ of the objects grew to masses of ${\rm m}\in(0.25,1)\,{\rm M}_\oplus$.

As seen in Figures \ref{Fig:eccentricity} and \ref{Fig:inclination}, the planetesimal disk spreads beyond the initial edge limit of 35 au. Under the assumption that the disk spread beyond 35 au, Neptune would not stop migrating at 30 au, as it would still have the ability to undergo planetesimal driven migration. However, the outer regions of the disk that spread viscously is not  dynamically cold, and could potentially violate the conditions necessary for planetesimal driven migration. Moreover, as depicted in Figure \ref{Fig:eccentricity} and Figure \ref{Fig:inclination}, large planetary embryos formed within the disk by T = 8Myr. Clearly, we do not see either of these features represented in the structure of the Kuiper Belt or Neptune's current orbit. In order to alleviate this discrepancy, Neptune would have had to have undergone rapid planetesimal driven migration. This would  eliminate the opportunity for large planetary embryos to form within the disk. Therefore, we conclude that the  instability must have occurred very early in the formation of the Solar System.

   

It is crucial to keep in mind that our results are contingent upon the assertion that the competition between viscous stirring and collisional damping prevents the disk's eccentricity and inclination distributions from widening too much. We verified that in the absence of damping, no collisional growth ensues in our calculation.  Because the effects of the disk's dynamical friction are implemented within our model through a fictitious damping acceleration, this effect is governed by our choice of a finite value for $\tau$. As already noted above, however, the $\tau\rightarrow\infty$ assumption leads to an unphysically over-excited disk, and therefore significantly underestimates the potential for collisional growth within the system. We further note that the scaling between $\langle e \rangle$ and $\tau$ is rather forgiving, such that choosing a somewhat higher value for the damping timescale in our model  should yield similar results. 

To verify this, we ran analogous simulations with 4 different damping timescales, $\tau$, and calculated the median eccentricity and inclination for all of the planetesimals after the disk reached an equilibrium state,  $\sim 1$Myr for all cases. The results of these experiments are shown in Table \ref{tab:tau}. Evidently, the median eccentricity and inclination do not sensitively depend on the magnitude of $\tau$. 

\begin{table}[]
    \centering
    \begin{tabular}{c|c|c}
    \hline
    \hline
          Damping Time & Median Eccentricity &Median Inclination \\
        \hline
         $2\times 10^4 \,{\rm yrs}$ &0.011 &0.20$^\circ$ \\
          $3\times 10^4 \,{\rm yrs}$ &0.013 &0.31$^\circ$ \\
          $5\times 10^4 \,{\rm yrs}$ &0.014 &0.31$^\circ$ \\
          $1\times 10^5 \,{\rm yrs}$ &0.018 &0.46$^\circ$ \\
    \hline
    \hline
    \end{tabular}
    \caption{Median eccentricity and inclination in the numerical simulations as a function of damping time, $\tau$.}
    \label{tab:tau}
\end{table}


\section{Discussion}\label{sec:discussion}

In this work, we explored the possibility of post-nebular oligarchic growth of protoplanetary embryos in the Solar System's primordial disk of debris. Overall, our numerical experiments indicate that it is plausible for $\sim 1{\rm M}_\oplus$ objects to emerge within the ancient trans-Neptunian region (i.e., $\sim15-35\,$au) on a $\sim10\,$Myr timescale, provided that the disk's velocity dispersion remains temperate. Accordingly, our numerical experiments indicate that if the planetesimal disk were to remain unperturbed for $\sim10-100\,$Myr, the present-day architecture of the Solar System likely would have been very different.

In virtually all respects of the physical setup, our simulations are similar to standard realizations of the Nice model  \citep{Levison2008,Levison2011,batygin011,Nesvorny2012}.  Previously published self-gravitating ${\rm N}-$body simulations of early dynamical instabilities in the Solar System had  interactions among super-particles that led to an unphysical degree of self-excitation within the planetesimal disk \citep{Ruiz2015,Fan2017}. In our simulations, we  introduced dissipation within the disk to approximately capture the effects of collisional damping. This effect maintains the planetesimals' eccentricities and inclinations on the order of a few percent and few degrees respectively, and facilitates accretion of protoplanetary embryos. Moreover, in this scenario,  gravitational scattering forces a substantial amount of icy debris to spread  beyond the present day orbit of Neptune.

Cumulatively, our results translate to an intriguing constraint on the instability timescale. That is, if the Solar System's transient phase of planet-planet scattering were to be delayed by hundreds of millions of years as envisioned in the original version of the Nice model \citep{Gomes2005}, the disk would have coalesced into planets, compromising its capacity to serve as the trigger mechanism for the instability. Consequently, our results suggest that the outward migration of Neptune would have started shortly after the dissipation of the protoplanetary nebula. This result is consistent with the conclusions of recent work that favored the early instability scenario \citep{Nesvorny2018b,Clement2018,Clement2019,deSousa2019,Nesvorny2021}.

There are numerous ways in which our study could be expanded upon. A limitation of our calculation is the relatively small number of particles in the simulation. Numerical methods such as those implemented in the \texttt{GENGA} code \citep{Grimm2014}, which uses GPUs to calculate gravitational interactions between bodies and produces results consistent with  \texttt{mercury6} \citep{1999MNRAS.304..793C}, could be used to expand these simulations, such as in \citet{Quarles2019}.  

 We have also ignored the initial size-distribution of particles, although  \citet{Kaib2021} found that in order to explain the existence of Pluto and Eris, the number of $\sim$Pluto-mass  bodies in the primordial Kuiper belt would have been as few as $\sim200$ and no greater than $\sim1000$. \citet{Shannon2018} provided similar limits on the number of primordial $\sim$Pluto-mass and  $\sim$Earth-mass protoplanets by considering the survival of ultra-wide binary TNOs, the Cold Classical Kuiper belt, and the resonant population. If these large planetesimals formed during the gaseous disk stage and were present in the primordial disk, this would likely speed up accretion such that Earth-mass objects would emerge on even shorter timescales via the same physical processes explored in our simulations. If instead these objects formed after the disk dispersal, then the  simulations described in the previous paragraph could test this hypothesis. For computational efficiency, our simulations used   accreting  embryos that are slightly larger than Pluto initially. We expect that  Pluto mass objects could have formed at some t $<$ 10 Myr based on our collisional growth calculations. 

In any case, considerations of growth in the disk provide an intriguing possibility to further constrain the instability driven scenario of the early evolution of the Solar System.



\acknowledgments
We thank the reviewer for insightful comments and constructive suggestions that helped the content of this manuscript. We thank Robyn Sanderson, Andrew Youdin, Adina Feinstein, Kaitlin Kratter, and Alessandro Morbidelli  for useful conversations.  K.B. is grateful to Caltech, and the David and Lucile Packard Foundation for their generous support. M. M. was funded by the Roy and Diana Vagelos Science Challenge Award. This research used the \texttt{numpy} \citep{Harris2020} and \texttt{matplotlib}  \citep{Hunter2007} packages in \texttt{python}.

\bibliography{sample63}{}
\bibliographystyle{aasjournal}



\end{document}